\documentclass[prl,twocolumn]{revtex4}
\usepackage{graphicx}% Include figure files
\usepackage{dcolumn}% Align table columns on decimal point
\usepackage{bm}% bold math
\usepackage{textcomp}
\usepackage{float}
\usepackage[colorlinks]{hyperref}
\usepackage{epstopdf}
\newcommand{\RR}{\right}
\newcommand{\LL}{\left}
\newcommand{\m}{\mathrm}
\newcommand{\eref}[1]{Eq.~(\ref{#1})}

\begin{document}

\title{Micromanipulation transfer of membrane resonators for circuit optomechanics}

\author{Maria Berdova$^{1,2,\dag}$}
\author{Sung Un Cho$^{1,\dag}$}
\thanks{Present address: Korea Research Institute of Standards and Science, Daejeon 305-340, Republic of Korea}
\author{Juha-Matti Pirkkalainen$^1$}
\author{Jaakko Sulkko$^1$}
\author{Xuefeng Song$^1$}
\author{Pertti J. Hakonen$^{1}$}
\author{Mika A. Sillanp\"a\"a$^{1,3}$}

\affiliation{$^1$O.~V.~Lounasmaa Laboratory, Aalto University, P.O. Box 15100, FI-00076 Aalto, Finland. \\$^2$Department of Materials Science and Engineering, Aalto University, P.O. Box 16200, FI-00076 Aalto, Finland \\
$^3$Department of Applied Physics, Aalto University, P.O. Box 11100, FI-00076 Aalto, Finland \\
$^\dag$these authors contributed equally to this work}

%\date{\today}% It is always \today, today, but any date may be explicitly specified

\begin{abstract}
A capacitive coupling between mechanical resonator and a microwave cavity enables readout and manipulation of the vibrations. We present a setup to carry out such experiments with aluminum membranes fabricated as stamps and transferred in place with micromanipulation. The membrane is held in place by van der Waals forces, and is supported by three microscopic points.  We measure the lowest mechanical modes, and conclude the membrane vibrates as an essentially free-free resonator. Sliding clamping conditions are identified via a softening Duffing nonlinearity. The method will enable reduction of clamping losses, while maintaining a narrow vacuum gap for strong capacitive coupling.
\end{abstract}

%\pacs{Valid PACS appear here}% PACS, the Physics and Astronomy Classification Scheme.

\maketitle

Micromechanical resonators are widely used everywhere in technology. A growing motivation for their basic studies is the fact that thus far micromechanical vibrations serve as the only model system for the quantum nature of motion. These works can take advantage of coupling the mechanical degree of freedom to electrical resonances in the microwave frequency regime. The electrical systems, often in the form of on-chip microwave resonators dubbed as cavities, act in a dual role of both detecting and manipulating the motion \cite{LehnertNphOpto08,SillanpaaCNEMS}. For studies near the quantum limit, the vibrating object should be strongly coupled to the electrical mode, which necessitates a narrow vacuum gap between a metallic resonator and a gate to maximize the capacitive coupling. Second, the mechanical quality factor should be as high as possible.

Clamped-clamped beam resonators are the most commonly used concept. The strongest capacitive coupling with beams is realized by Focused Ion Beam cutting of a 10 nm vacuum gap \cite{SulkkoFIB,MechAmpPaper}. Recently, the best results in studies near the quantum limit have been obtained not with beam geometry but using a drumhead, which maximizes the participation of the volume to the movable capacitance \cite{Teufel2011a}.

%%%%%%%%%%%%%%%%%%%%%%%%%%%%%%%%%%%%%%%%%%%%%%%%%
\begin{figure}[tH]
\centering
\includegraphics[width=0.71\linewidth]{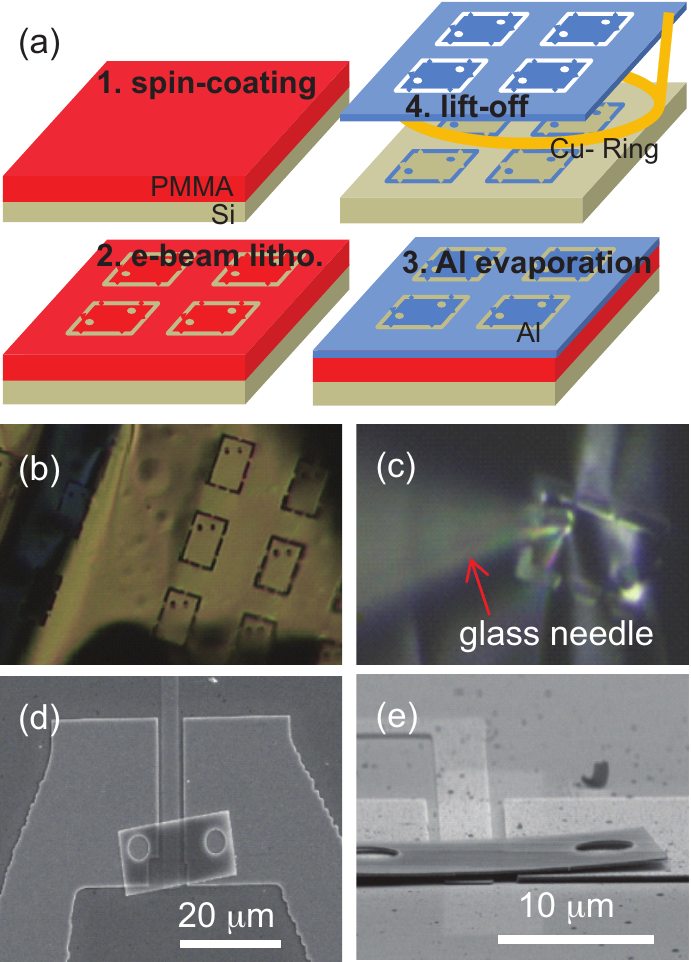}
\caption{(a) Fabrication procedure for harvesting a mm-sized aluminum sheet patterned full of stamps which will be used as the vibrating objects. (b) Optical micrograph of the Al sheets (150 nm thick) after the lift-off process, and (c) of peeling off an individual Al stamp with a glass needle. (d), (e) Scanning electron micrographs of an Al stamp stacked over a trench, top (d) and side (e) view.}\label{fab}
\end{figure}
%%%%%%%%%%%%%%%%%%%%%%%%%%%%%%%%%%%%%%%%%%%%%%%%%

In order to realize the vibrating objects, the suspended structures are defined through a selective chemical etching of a sacrificial layer. However, etching is limited to specific material because the requirement of chemical compatibility with the rest of the structure. Here, we introduce a scheme to realize a membrane or bridge resonator by assembling a metallic "stamp" into a measurement platform patterned on a Si chip. The method can flexibly adapt various conducting or non-conducting \cite{Weig2011} materials, as well as be combined with numerous other processes because etching compatibility is irrelevant. Our approach is a promising way towards achieving strong capacitive coupling, and reduction of mechanical losses by minimizing acoustic radiation into the supports \cite{Nguyen2000,AspelmeyerTunn} as clamps are essentially eliminated in our design.

The fabrication process is related to our previous work on micromanipulation transfer of graphene mechanical resonators \cite{GrapheneStamp}. The whole set of the process, see Fig.~1, is divided into three steps; aluminum (Al) sheet fabrication, building up the metallic gate and support structure for coupling to the microwave resonator, and piecing these parts together.

First, stamp shape structures are patterned by regular electron-beam lithography through a PMMA mask, followed by evaporation of 150 nm of Al. The individual stamps used are $22$ $\mu$m long and $11$ $\mu$m wide. We believe smaller or larger stamps can be assembled similarly, however, care has to be taken in order to avoid deformation in stamps larger than presently used. The stamp has a couple of $4$ $\mu$m diameter holes for picking it up with a glass needle. In order to make a fine glass needle to fit in the hole, glass capillary is gravitationally elongated and snapped under local heating applied midpoint of the capillary. The Al sheet filled with stamps becomes released in a lift-off in acetone, and it is scooped up with a copper ring, Fig.~1 (a). Finally, a single Al stamp is torn from the sheet with glass needle and stacked on the measurement platform using micro-manipulation, as shown in Fig.~1 (c).

The platform structure where the Al stamp is stacked on, is also patterned by e-beam lithography on a Si chip. The first lithography patterns the 50 nm thick gate serving as the counter electrode for the movable capacitance $C_g(x)$ as well as connecting to the microwave cavity. The electrodes, 250 nm thick, connected to electric ground on each side of the gate line physically support the stamp. The width of the gate line is $4$ $\mu$m, and the ground electrodes are separated from each other by $5$ $\mu$m. Images of the assembled structure are shown in Fig.~1 (d), (e). In a highly tilted micrograph, we also measure the vacuum gap of $\sim 240$ nm.

As can be seen in the tilted micrograph of Fig.~1 (e), the assembled stamp retains its slightly curved shape during assembly, similarly to a piece of aluminum foil dropped on a surface. This indicates the bending rigidity dominates over surface adhesion, and hence we believe the stamp is supported by three points of size down to atomic scale. As we argue below, the stamp can slide along the support points when moderately pulled by a gate voltage, as displayed in Fig.~2 (a).

For the measurement, we use the "circuit optomechanics" architecture, which borrows the concepts from optical interferometers \cite{KippenbergReview}. Briefly, a microwave pump tone  (frequency $f_P \sim 4$ GHz) injected into the microwave cavity is amplitude- and phase modulated by the time-dependent capacitance. The gate is also supplied with a gate frequency ($f_g$) on top of a static dc voltage for actuating mechanical motion. The driving force will be $F = V_{\m{dc}} V_{\m{ac}} \LL( d C_g/dx \RR)$. The dc also has the function of pulling the membrane. The voltages are combined at room temperature, see Fig.~2 (b). The sidebands at frequencies $f_{\pm} = f_P \pm f_g$ are detected by a spectrum analyzer.

Our experiment is conducted at liquid helium temperature $\sim 4$ K, and at a helium exchange gas pressure of $\sim 10^{-3}$ mbar. While the actuation frequency $f_g$ is swept, we observe peaks in the sideband voltage which are identified as the mechanical eigenmodes being excited, see Fig.~2 (c). The angular frequency of the mode under consideration is marked as $\omega_m = 2 \pi f_m$.

%%%%%%%%%%%%%%%%%%%%%%%%%%%%%%%%%%%%%%%%%%%%%%%%%
\begin{figure}[tH]
\centering
\includegraphics[width=0.98\linewidth]{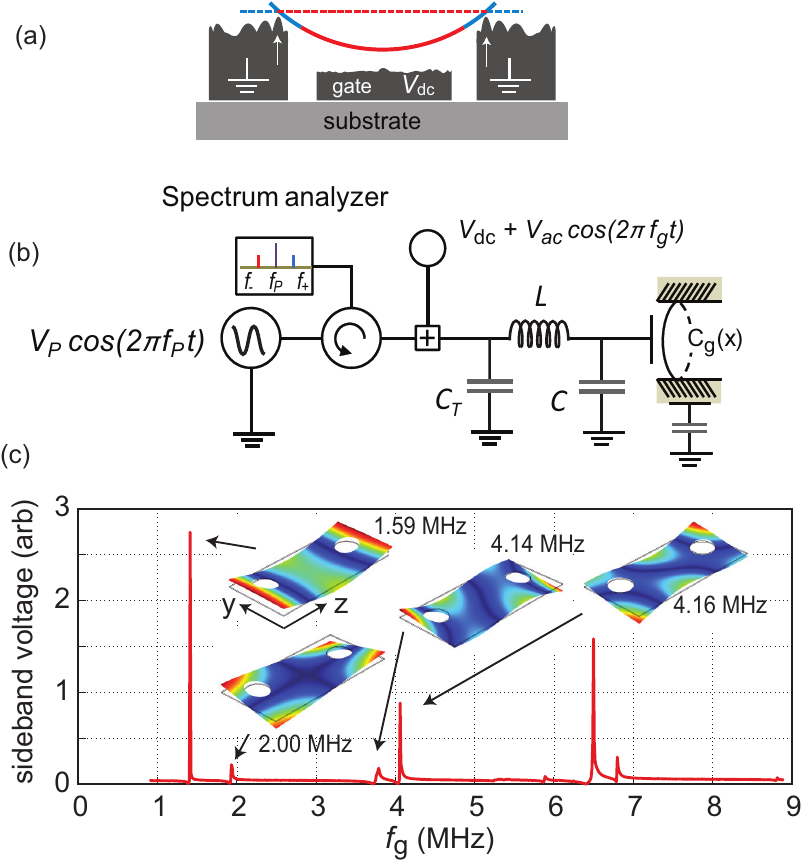}
\caption{ (a) Cross-section of the stamp sitting on the platform, illustrating the sliding clamping conditions. The membrane is pinned from three points defined by the surface roughness (arrows, two shown), and is originally straight (dashed line) but it can slide from the supports when pulled towards the gate by dc voltage (full line). For clarity, the area outside the original clamping points is shown in blue, and the rest in red. (b) Schematic diagram of measurement set-up for detection of the membrane eigenmodes from the modulated probe microwave. The $LC$ tank circuit ("cavity") has a stray capacitance $C \sim 0.3$ pF, and a tuning capacitance $C_T \sim 3$ pF used to enhance the external electrical $Q$-value. (c) Result of scanning the gate excitation frequency, revealing some of the lowest modes visible as peaks. In the insets, the four lowest modes are indicated with a simulation of free-free membrane mode shapes together with their simulated frequencies. }\label{schema}
\end{figure}
%%%%%%%%%%%%%%%%%%%%%%%%%%%%%%%%%%%%%%%%%%%%%%%%%

In order to understand the mode structure, we made Finite Element simulations of the stamp structure. The simplest model is that of a free-free membrane, that is, a seemingly unphysical levitated object. The lowest mode shape depends up to a good approximation only on the coordinate $z$ along the length beam, and is approximated as
\begin{equation}
 X(z,y) = X_0 \LL( 2 \sin \LL(\pi \frac{z}{L} \RR)-1 \RR) \,.
\label{mode}
\end{equation}
We normalize the displacement to the maximum deflection amplitude $X_0$.

The simulated four lowest eigenmodes, displayed in Fig.~2 (c) and connected to the measured peaks, match well with the measured frequencies. The height of each peak is proportional to the electromechanical coupling which here comes from $d C_g(x)/dx$ integrated over the gate area in the middle of the membrane. Although we have not made a quantitative comparison, this quantity can be seen, based on the mode shapes, to be the largest for the lowest mode, and nearly vanishing for the second, in agreement with the  measured relative peak heights. We can also simulate the stamp being supported by different types of clamp configurations. With a generic clamped membrane, we cannot reproduce the measured mode frequencies. Therefore, we conclude the membrane is essentially vibrating as if it were levitated.

Further evidence of the "sliding pinned" boundary conditions is obtained by studying the nonlinear (Duffing) regime of vibrations \cite{NonLinOsc,DuffingSign,SoftDuff,Buks06APL,CollinDuff2012}. An intriguing feature is the softening Duffing nonlinearity \cite{DuffingSign,SoftDuff} observed for the first mode (Fig.~3 (a)), which is also readily explained. Generally, lengthening of a clamped beam under deformation induces tension, which further stiffens the spring. Hence, a clamped beam experiences a stiffening Duffing (nonlinearity) as the vibration amplitude grows. Another source of nonlinearity is due the electrostatic potential between the beam and gate. The attractive force becomes the stronger the larger the deformation is, and hence electric nonlinearity gives rise to softening Duffing. The overall sign depends then on the ratio of these two. In the present case, supposing sliding supports, the mechanical nonlinearity would be absent, and thus is a tempting explanation for the softening Duffing behavior.

We will first carry out full analysis including a possibility for the mechanical nonlinearity. The elastic energy is from basic elasticity theory \cite{Blanter2003PRB}
\begin{equation}
E_{\mathrm{bend}}=  \frac{EI}{2}\int_0^L\left[\frac{d^2 X}{dz^2}\right]^2 dz +\frac{E A}{8L}\left(\int_0^L\left[\frac{d X}{dz}\right]^2 dz\right)^2
\label{elastinen}
\end{equation}
Here, $X(z)$ is the mode shape along the length $z$ of the beam. $I$ is the moment of inertia, $L$ and $A$ are the length and cross-section area of the beam, respectively.

%%%%%%%%%%%%%%%%%%%%%%%%%%%%%%%%%%%%%%%%%%%%%%%%%
\begin{figure}[tH]
\centering
\includegraphics[width=0.86\linewidth]{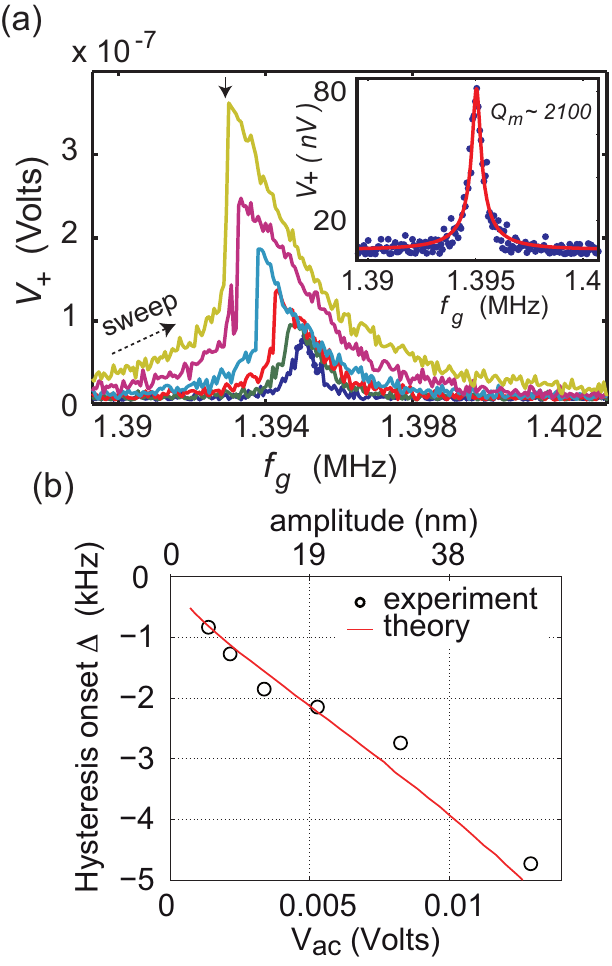}
\caption{(a) Measurement of the first eigenmode which resembles a free-free membrane vibrations. The driving voltage $V_{\m{ac}}$ is increased from bottom to top, from 1 mV (blue) to 14 mV (yellow) with 5 dB intervals. Temperature was $T = 4.2$ K, $V_{\m{dc}} = 1$ V, $V_{\m{P}} \sim 2$ mV. The sweep direction was from left to right. The hysteresis jump is marked by a vertical arrow. The inset shows the response in the linear regime, together with a Lorentzian fit. (b) Position of the hysteresis jump with respect to bare resonance frequency, as a function of drive amplitude, together with a theoretical prediction from a simple driven Duffing oscillator.}\label{schema}
\end{figure}
%%%%%%%%%%%%%%%%%%%%%%%%%%%%%%%%%%%%%%%%%%%%%%%%%

Let us consider the mode shape \eref{mode} of the lowest mode but with pinned at the nodal points. The mechanical nonlinearity, the second term in  \eref{elastinen} about 50 \% of the full value $\frac{\pi^4  E A}{2 L^3} X_0^4$ when integrated between the nodal points instead of full beam length. The electrical contribution to the potential energy is
$V= -\frac{1}{2} C_g(X_0) V_{\m{dc}}^2$.

We expand the total displacement $X_0 = X_{00} + x_0$ around an equilibrium position and small fluctuations $x_0$. This gives the total potential
%
%V= -\frac{1}{2} C(Y_0) V^2 =  \frac{1}{2} V''( Y_{00}) Y_0^2 +\frac{1}{6} E_p^{(3)}( Y_{00}) Y_0^3 +\frac{1}{24} V^{(4)}( Y_{00}) Y_0^4 + ...
%
\begin{equation}
V= \frac{1}{2} M \omega_{\m{eff}}^2 x^2_0 + \frac{1}{3} K_2 x_0^3 + \frac{1}{4}K_3 x_0^4
\label{duffing}
\end{equation}
Here, $M$ and $ \omega_{\m{eff}}$ are the effective mass and the effective frequency of linear oscillations, respectively. The nonlinear terms have opposite-sign contributions from mechanical and electrical nonlinearity.
%$$
%K_2 = \frac{1}{6} \LL[ 24 \delta Y_{00} - \frac{3V^2}{(d-Y_{00})^4}\RR] 
%K_3 = \frac{1}{24} \LL[ 24 \delta - \frac{12V^2}{(d-Y_{00})^5}\RR]
%$$
%The Duffing equation including both third and fourth order (in energy) nonlinearities is
%
%\begin{equation}
 %\ddot{X}_0  + \gamma \dot{X}_0  + \omega_0^2 X_0+K_2 X_0^2+K_3X_0^3= F/M \cos(\omega_g t)
%\label{duffing}
%\end{equation}
%
Harmonic-balance solution \cite{NonLinOsc} to the corresponding Duffing equation gives an effective Duffing term contributed by both the third and fourth order nonlinearities \cite{nonlinNR,Khan2013}:
\begin{equation}
K_3^* = K_3 -  \frac{10}{9} \frac{K_2^2 }{ M \omega_m^2}
\label{duffeff}
\end{equation}
One possible benchmark is the location of the hysteresis jump $\Delta$ nearest to the linear-regime frequency. In the limit that the mechanical damping rate $f_m/Q_m \ll \Delta$ we obtain
\begin{equation}
\Delta= \m{sign} \LL({K_3^*}\RR) \frac{3 \sqrt[3]{3/2 K_3^* F^2}}{4 M \omega_m}
\label{jump}
\end{equation}
We can also verify numerically that this holds reasonably well even if the amplitude just exceeds the hysteresis threshold $\Delta$. Equation (\ref{jump}) is compared to the experiment in Fig.~3 (b), using the attenuation in the gate drive line as a fitting parameter. We have cut off the theory curve below the mentioned rough validity region $f_m/Q_m \sim \Delta$. With any reasonable parameters, one cannot obtain a softening nonlinearity (negative $K_3^*$ in \eref{duffeff}). We thus conclude that sliding clamping conditions are a good description of the stamp dynamics. However, the present experiment does not allow to separate the clamping loss contribution to dissipation, since the measured values  $Q_m \sim 2 \times 10^3$ are typical of those of Al resonators at 4 K. These have internally limited losses based on the strong temperature dependence of $Q_m$ in Al devices below about 1 K  \cite{SulkkoFIB}, whereas clamping loss is expected to be nearly temperature-independent.

In conclusion, we have demonstrated a circuit optomechanics experiment where the mechanical resonator is a nearly free-free membrane. This is obtained by micromanipulation transfer of an aluminum "stamp" on top of a measurement platform. This opens up perspectives for realizing suspended structures having low losses from supporting clamps, and without the need of caring for etching compatibility of a sacrificial layer.

\begin{acknowledgments}
This work was financially supported and by the Academy of Finland though its LTQ CoE grant (project no. 250280) and the projects 120058, 110058 and 259912, by the Cryohall infrastructure, by the V\"ais\"al\"a Foundation, by the European Research Council (grant No.~FP7-240387), and by the EU 7th Framework Programme (Grant No.~228464 Microkelvin).
\end{acknowledgments}

%\bibliographystyle{apsrev}
%\bibliography{/Users/masillan/Documents/latex/mikabib}

%\begin{thebibliography}{99}
%\bibitem{Regal2008} Regal, J. D.; Teufel, C. A.; Lehnert, K. W. \emph{Nature Physics} \textbf{2008}, 4, 555-560.
%\bibitem{Sillanpaa2009}Sillanp\"a\"a, M. A.; Sarkar, J.; Sulkko, J.; Muhonen, J.; Hakonen, P. J. \emph{Appl. Phys. Lett.} \textbf{2009}, 95, 011909-011911.

%\end{thebibliography}
\end{document}